\documentclass{article}
\usepackage[utf8]{inputenc}
\usepackage{braket}
\usepackage{authblk}
\usepackage{hyperref}
\usepackage{bm}
\usepackage{amssymb}

\title {The Neverending Story of the Eternal Wormhole and the Noisy Sycamore}
\author{Galina Weinstein}
\affil{\normalsize Reichman University, The Efi Arazi School of Computer Science, Herzliya; University of Haifa, The Department of Philosophy, Haifa, Israel.} 

\begin{document}

\maketitle

\begin{abstract} 
There has been a great buzz surrounding Daniel Jafferis et al.'s latest \emph{Nature} paper, ”Traversable wormhole dynamics on a quantum processor”. The \emph{Nature} paper discusses an experiment in which Google's Sycamore quantum processor is used to simulate a sparse $N = 7$ SYK model with $5$ terms (a learned Hamiltonian). The \emph{Nature} paper shows that the learned Hamiltonian preserves the key gravitational characteristics of an $N=10$ SYK model with $210$ terms and is sufficient to produce a traversable wormhole behavior. I will examine the experiment and discuss some philosophical challenges concerning the experiment in memory of Ian Hacking. Recently, Norman Yao and two graduate students discovered multiple flaws in Jafferis et al.'s learned Hamiltonian and uploaded a comment on the \emph{Nature} paper. As expected, Jafferis and his team found a simple way to clarify the misunderstanding. They found a physical justification that allowed them to avoid the problem. In this paper, I elucidate the main arguments Yao and his students raised and the way Jafferis et al. found to save their learned Hamiltonian. I will end this paper with a philosophical comment on this recent development in the context of the learned Hamiltonian. 

\end{abstract}

\section{Introduction}

In 1993 Charles H. Bennet and other IBM Thomas J. Watson Research Center authors invented quantum teleportation. They began their paper, “Teleporting an unknown quantum state via dual classical and Einstein-Podolsky-Rosen Channels”, in which they suggested teleportation by saying: 
“The existence of long-range correlations between Einstein-Podolsky-Rosen (EPR) pairs of particles raises the question of their use for information transfer". 
In his \emph{Autobiographical notes}, "Einstein himself used the word 'telepathically' in this context. It is known that instantaneous information transfer is impossible. Here, we show that EPR correlations can assist in the 'teleportation' of an intact quantum state from one place to another by a sender who knows neither the state to be teleported nor the location of the intended receiver”. A teleportation protocol is then described between Alice and Bob. Accurate teleportation is achieved only by Alice sending the outcome information as classical bits to Bob \cite{Bennet}.    

In 2017 Leonard Susskind and Ying Zhao suggested that one could teleport a unit of quantum information (a qubit $|\psi\rangle$) between the two sides of a microscopic Einstein Rosen (ER) bridge, connecting an entangled EPR pair. The conditions for teleportation render the ER bridge traversable so that a qubit entering one end of the ER bridge will, after a suitable time, appear at the other end.

Susskind and Zhao explain that although we cannot send information faster than light, a qubit is teleported through the microscopic so-called "baby" ER bridge, connecting the entangled EPR pair. However, the ER bridge is too small for such a microscopic system to have a classical space-time geometry. 

On the other hand, on the macroscopic level, according to the ER = EPR relation, an ER bridge between two black holes (joined by a throat) implies that they are entangled, and entanglement between two black holes implies that an ER bridge connects them. We can, therefore, hypothetically apply the same reasoning of teleportation as communication of quantum information through space-time wormholes when the entangled EPR systems are a pair of entangled macroscopic black holes. Then we may hope to follow the geometry of the wormhole as the teleported system passes through it \cite{Susskind}.

Yet Susskind and Zhao know, “The operations we’ve described would be very hard, if not impossible, for real black holes, but we can imagine laboratory settings where similar things may be possible. Suppose that in our lab, we have two non-interacting large shells of matter, each of which has been engineered by condensed matter physicists to support conformal field theories of the kind that admit gravitational duals. If we make the two shells out of entangled matter, we can produce the shells in the thermofield double state for some temperature” \cite{Susskind}.     
Would there be a hidden little wormhole connecting the shells? Assuming a model in which the entire laboratory is embedded in a world that satisfies ER=EPR, then in this model laboratory, there would be such a wormhole. But we are not yet speaking of our real world - I will discuss this matter further in section \ref{8}.   

To confirm a hidden wormhole, Alice and Bob can merge themselves with the matter forming the shells and eventually be scrambled into the conformal field theory (CFT) thermal state - I will discuss scrambling in section \ref{1}. They would fall into their respective black holes in the dual gravitational picture. If conditions were right, they would meet before being destroyed at the singularity; all this takes place outside ordinary space. Unfortunately, they could not inform the exterior world that the wormhole was real or that they successfully met.
However, quantum teleportation allows Alice and Bob to confirm the existence of the wormhole without jumping in. Alice may convince Tom to jump into the wormhole, which she and Bob can render traversable by introducing a temporary coupling. When Tom emerges out of the degrees of freedom of Bob’s shell, he will recall everything he encountered and can confirm that he did traverse the wormhole \cite{Maldacena2}.

Susskind and Zhao’s paper extends Susskind’s 2016 idea of quantum teleportation through Schwarzschild wormholes (ER bridges). Teleportation through a Schwarzschild wormhole violates classical general relativity. Schwarzschild wormholes are non-traversable because they do not allow us to send a signal from one asymptotic region to another when suitable positive energy conditions are obeyed. According to the ER = EPR relation, an ER bridge between two black holes is created by EPR-like correlations between the microstates of the two black holes \cite{Susskind1}. 

In the same year, Ping Gao, Daniel Louis Jafferis, and Aron C. Wall showed that the ER = EPR relation allows the ER bridge to be traversable. Gao, Jafferis, and Wall suggested that entanglement alone cannot transmit information for the ER bridge to become traversable. We need quantum teleportation: Alice and Bob share an entangled pair of qubits $A$ and $B$. Alice is given a qubit $Q$ to be transmitted to Bob. Alice performs a joint measurement of $Q$ and $A$ and sends the result to Bob as classical bits of information. Bob will perform a unitary operation on his qubit $B$, which depends on Alice’s results and gets qubit $Q$. “Of course, in the limit that Alice’s measurement is essentially instantaneous and classical, the traversable window will be very small […] – just enough to let the single qubit $Q$ pass through. Therefore, we propose that the gravitational dual description of quantum teleportation understood as a dynamical process is that the qubit passes through the wormhole of the entangled pair, $A$ and $B$, which has been rendered traversable by the required interaction” \cite{Gao1}.

In the traversable wormhole picture, the qubit "may be identified with the system that falls into the black hole from the left and gets scrambled, the auxiliary entangled system is on the right, and the boundary interaction somehow triggers the appropriate quantum computation to make the qubit reappear again, after a time of order the scrambling time” \cite{Gao1}. I will thoroughly discuss this protocol in sections \ref{1}, \ref{2}, \ref{3}, and \ref{5}. 

Gao, Jafferis, and Wall conclude their article by saying that their example "provides a way to operationally verify a salient feature of ER=EPR that observers from opposite sides of an entangled pair of systems may meet in the connected interior": If, after the observers jump into their respective black holes, a coupling is activated, then the ER bridge can be rendered traversable, and the meeting inside may be seen from the boundary \cite{Gao1}.
The above traversable wormhole does not require energy matter that violates the averaged null energy condition (ANEC) because the negative energy-matter in the ER = EPR configuration is similar to the Casimir effect (providing negative energy density). Further, any infinite null geodesic which makes it through the wormhole must be chronal, i.e., the wormhole does not violate Hawking’s chronology protection conjecture. 

In 2017 Susskind shed some light on the traversable wormhole \cite{Susskind1}: “It is especially interesting that if Tom encounters objects during his passage from Alice’s side to Bob’s side, his experiences may be recorded in his memory. This would allow him to report the conditions in the wormhole to laboratory observers. Quantum teleportation through the wormhole is a real game-changer; it provides a direct way to observe the interior geometry of a wormhole. One can no longer claim that life behind the horizon is unphysical, meaningless, unobservable, or scientifically unfalsifiable”. Susskind pointed out that quantum teleportation of a single qubit through a channel of a single Bell pair was already an experimental reality. ER=EPR would allow it to be interpreted as teleportation through a Planckian wormhole. He claimed this would open a possibility for seeing quantum gravity in a lab equipped with quantum computers. 

I will discuss quantum gravity in the lab in section \ref{9}.

Susskind asked in 2017 \cite{Susskind1}: “Can laboratory experiments of this type be carried out? I don’t see why not. Instead of shells supporting conformal field theories, a more practical alternative might be quantum computers simulating the CFTs. Entangling two identical quantum computers into a thermofield double state should be feasible. To teleport a genuine sentient Tom through the wormhole would require many qubits. Still, with a few hundred logical qubits, one can teleport a register composed of say, ten qubits - enough for a primitive memory". The operations needed for the kind of laboratory teleportation offered above, said Susskind in 2017, “are fairly complex (in the computational sense) and are therefore difficult”, but Susskind did not see “anything forbidding them once quantum computers become available”.

A version of Susskind's suggestion was carried out on Google's Sycamore processor by Jafferis and a team of researchers, among which are Alexander Zlokapa and Maria Spiropulu and the results were published in \emph{Nature}\cite{Jafferis}. This paper aims to critically discuss the experiment. 

Recently, Norman Yao, along with two colleagues Bryce Kobrin and Thomas Schuster, discovered multiple flaws in Jafferis et al.'s learned Hamiltonian [see equation (\ref{Eq2}) in section \ref{3}] and published a comment on the \emph{Nature} paper. In other words, Korbin et al. discovered multiple problems in the model of quantum teleportation. As expected, Jafferis and his team found a simple way to clarify the misunderstanding and a physical justification that allowed them to avoid the problem. I will discuss Korbin et al.'s and Jafferis et al.'s comments in section \ref{10}.  
I will further discuss philosophical challenges concerning an experiment in memory of Ian Hacking in section \ref{8}.   

\section{Quantum chaos and scrambling} \label{1}

Let us begin with the quantum \emph{butterfly effect}, essential for understanding the experiment. The butterfly effect implies scrambling \cite{Hosur}. \emph{Quantum scrambling} is the quantum analog of chaotic dynamics in classical systems. Scrambling describes many-body dynamics which, though ultimately unitary, scatter initially localized quantum information across all of the system’s available degrees of freedom. Black holes are the fastest scramblers in the universe and are the most chaotic bodies in the cosmos \cite{Sekino}; \cite{Blok}. 

More specifically, quantum information in a small local space area spreads out, and we must search a large region to recover the information. This is the scrambling of the quantum information while the system evolves. Heisenberg's operators evolve in a way that reminds the chaotic butterfly effect, i.e., they were first local and are now spread over many regions in space. This is the butterfly effect in quantum systems. 

The models for the onset and dynamics of quantum chaos are called the \emph{Sachdev-Ye-Kitaev} (SYK) models. The SYK model of a many-body quantum system was first proposed in condensed matter physics. It is called after Subir Sachdev, Jinwu Ye, and Alexei Kitaev, who later modified it. It was found that the SYK model also has applications in other domains, such as AdS/CFT correspondence. 

It should be stressed that what is meant by black holes here and after that is \emph{eternal black holes} or \emph{anti-de Sitter space} (AdS) black holes. And we are also speaking of \emph{eternal wormholes}. The main feature of black holes in AdS is that the boundary reflects their Hawking radiation. This is unimportant for small enough black holes since the entire black hole could evaporate before the radiation reaches the boundary. Still, as the Schwarzschild radius of the black hole approaches the AdS radius, we eventually reach a point where the radiation is being reflected into the black hole as fast as it is being emitted. The black hole never evaporates at this point, so large enough black holes in AdS are therefore eternal \cite{Harlow}.
If two eternal black holes are close, this would give rise to an interaction between them connecting the two separated boundaries of AdS2. The interaction can produce a throat that connects the two sides of the black hole so that the final geometry is a horizonless \emph{eternal traversable wormhole}. 
The interaction creates \emph{negative energy} in bulk. Imagine a particle that starts from one boundary, living in the first system; it then moves to the other side, living in the second system. "This is somewhat reminiscent of two weakly coupled oscillators where the excitation moves from one oscillator to the other, except that it does this through the bulk geometry". The same property holds in the two copies of SYK models coupled by simple interaction. "The dynamics of the two coupled SYK systems looks like that of a traversable wormhole" \cite{Maldacena1}.

What this means is that the \emph{eternal wormhole} is dual to two copies of the original \emph{conformal field theory} (CFT) in the \emph{thermofield double} (TFD) state. The eternal black hole's left and right external bulk regions are joined through a wormhole and are thus dual to the TFD state \cite{Maldacena1}. The TFD state is an entangled pure state between two identical copies of the quantum system (CFT): 

\begin{equation} \label{eq1}
\frac{1}{\sqrt{Z}_{\beta}}\ket{TFD_{\beta}} = e^{-\beta(H_L+H_R)} \ket{n}_{L}\otimes\ket{n}_{R}. 	  
\end{equation}

Tracing over one of the two copies of the system, $H_L$ (the SYK Hamiltonian applied to the left system) or $H_R$ (the SYK Hamiltonian applied to the right system), gives a thermal state with Majorana fermions (at temperature $T=1/\beta$). The state on just side $L$ or just side $R$ is thermal. The $\ket{n}_{L, R}$ are the energy eigenstates of $H_L$ and $H_R$ acting on Hilbert spaces. 

The SYK models lead to scrambling and spreading information among the quantum many-body system. But, as said before, the SYK models possess gravity duals. They are also a paradigm for quantum holographic matter and the gravitational interpretation through the holographic principle or duality, the AdS/CFT correspondence or gauge/gravity duality, the equivalence between two descriptions of the same system: a gravitational theory in higher $(d + 1)$ dimensions, on the one hand, and quantum field theory (that does not contain gravity) in lower $d$ dimensions, on the other.
The AdS description is often called the \emph{bulk} theory, while the CFT nongravitational dual is called the \emph{boundary} theory. In AdS/CFT, the two theories are equivalent, and we should use either one description or the other but not both \cite{Harlow}. 
The AdS/CFT duality is a correspondence between gravitational systems in AdS space-times and nongravitational quantum CFTs. This duality has its roots in the notion of duality dating back to the early years of quantum physics: the wave–particle duality. A CFT in the thermofield double state of Eq. \ref{eq1} would be dual to the two-sided eternal black hole described above.

When looked at differently, there are chaotic quantum systems that some quantum space-time with gravity can equivalently describe. The main argument of theoreticians suggesting experiments called \emph{quantum gravity in the lab} is that experiments performed on a nongravitational boundary system can detect phenomena characteristic of a gravitational dual. Such experiments would enable us to search for signs of a bulk-gravitation dual for nongravitational quantum systems \cite{Brown2}, \cite{Nezami2}. 

An SYK model becomes extremely chaotic at the very beginning of its development. In the SYK model, the \emph{out-of-time-order correlation} (OTOC) functions are used to diagnose quantum chaos and measure the growth of operators in space, unitarily evolving (in the Heisenberg interpretation of quantum mechanics) as a function of time. With chaotic time evolution, the butterfly effect will cause most of the OTOC functions in the average to decay exponentially \cite{Hosur}.

In the semi-classical limit (in quantum systems with many degrees of freedom), this scrambling of information and operator growth due to chaotic behavior is exponential. It is measured using the quantum Lyapunov exponent. But unlike the classical Lyapunov exponent, a bound exists on the quantum Lyapunov exponent. This is additionally measured by the \emph{butterfly velocity}, the equivalent measure of the classical chaotic butterfly effect. The quantum Lyapunov exponent and the scrambling rate are the ones that characterize the beginning and appearance of quantum chaos in this system \cite{Maldacena}. 

It should be noted that an interesting characteristic of the SYK model, which is related to the quantum Lyapunov exponent and the OTOC, is that the model exhibits maximally chaotic behavior. Like eternal black holes, the SYK model is a very fast scrambler of information. 
Another important quantity called \emph{Loschmidt echo} is intimately tied to quantum chaos. The echo is the probability that the chaotic system would return to its initial state. 

As said above, we characterize quantum scrambling and quantum chaos by measuring the OTOC function. However, OTOCs do not generally discriminate between \emph{quantum scrambling} and the effects of both ordinary \emph{quantum decoherence} and experimental noise: quantum scrambling and classical noise lead the OTOC to decay exponentially with time. It is a major problem if quantum scrambling is indistinguishable from quantum decoherence and noise, where the information in a system is lost to the environment \cite{Landsman},\cite{Yoshida}. 

Isolated systems are idealized models, but unfortunately, realistic systems are open systems that interact with the environment. Suppose there is a system of $n$ qubits. This system is not isolated and closed. The $n$ qubits interact with many interfering particles in the complex environment. It is almost impossible to follow the dynamics of each particle, so what we have here is a system that is a many-body system and decoherence induced by the environment.
As the system evolves, the $n$ qubits get entangled with the many-body system of the environment, and there are more disturbances, perturbations, and degrees of freedom. Decoherence happens naturally to quantum computers since qubits can’t be perfectly isolated from the environment.

It was found that a quantum teleportation protocol enables one to differentiate between scrambling and decoherence. Thus, teleportation can verify scrambling behavior even in decoherence and experimental imperfection \cite{Blok};\cite{Yoshida}.

\section{SYK models and holography} \label{2}

The SYK Hamiltonian is a model for quantum chaos and holography. That is, there is correspondence between the SYK model and scrambling/quantum chaotic behavior on the one hand and eternal black holes on the other. This dual possibility led a team of researchers to realize they could create a teleportation model through a traversable wormhole. They discovered that a process called \emph{unscrambling} comes after scrambling in a wormhole. The discovery of scrambling followed by unscrambling has boosted the possibility of realizing a quantum mechanism called \emph{size winding} in the lab. This process completely goes against everything we know from classical chaos and irreversibility. The size-winding mechanism is reminiscent of \emph{Poincare's Recurrence Theorem} of classical physics. But in the dual gravitational interpretation, size-winding leads to the interesting conclusion that a particle can pass through the holographic wormhole. 

The protocol, described in section \ref{5}, is the following: the information is scrambled on the left side of the wormhole. Since the two sides, right and left of the wormhole, are connected (coupled), the information, i.e., qubits, is unscrambled and pops out on the right side. Two essential things enable traversability: the two sides of the wormhole must be entangled before sending the information, and the two sides must be coupled after sending the message. More specifically, one inserts a qubit (the message) on the system's left side ($L$ subsystem), and it evolves in time. The qubit is entangled with one of the qubits on the $L$ subsystem. It means the qubit spreads among the $n$ qubits (a small number of qubits) on the left side and in all subsystem parts. After a certain time, the qubit is entangled with the qubits of the $L$ subsystem. But then the qubit suddenly reappears, is unscrambled, and recoheres on the other side, the right side ($R$), very far from the $L$ side, where it was scrambled. 
Something caused the original qubit, which entered on the far-left side, to suddenly be focused on the far-right side at a future time, even though it was completely mixed up on the left side.
The above mechanism is called \emph{teleportation-by-size}. 

So Adam Brown, Susskind, and their team of researchers pondered the possibilities: one way to describe this phenomenon is to brute force using the Schrödinger equation. If we imagine that the two sides of the system represent two sides of the eternal black holes ($L$ and $R$) connected by a wormhole, then the explanation for the phenomenon is \emph{simpler}. A teleported message is sent through an emergent wormhole: it is injected into $L$ and arrives at $R$ later due to a coupling operator. Hence, this phenomenon has a simpler explanation in the setting of holographic quantum gravity. Teleportation is possible despite the chaos because the system behaves orderly. It is bizarre from the quantum mechanical point of view, says the researchers, but what makes things less weird is that it may be explained or interpreted using the paradigm of quantum gravity and the holographic principle: a traversable wormhole protocol is equivalent to the above quantum information protocol \cite{Brown}, \cite{Brown2}. 

Researchers decided to test their hypothesis of many-body teleportation - the dynamics of a traversable wormhole through which a qubit can pass - by simulating the SYK model of $N$ Majorana fermions on a Google quantum device. It was suggested that realizing the holographic SYK model on the Google Sycamore chip might open a window to understanding the quantum gravity of holographic traversable wormholes. 

\section{Perfect size winding and the learned Hamiltonian} \label{3}

Brown, Susskind, and their team suggest an Ansatz for the traversable wormhole: \emph{size winding}. The researchers first describe size-winding from the boundary point of view and then apply it to the traversable wormholes (in bulk). 

In the Heisenberg picture, near the \emph{scrambling time} for the SYK model, a thermal operator $P$ is inserted at a negative time into the left boundary (the left side $L$). Recall that the growth of the size of an operator is a basic manifestation of quantum chaos and complexity of the system. The operator-size distribution is winding in the clockwise direction. A coupling is applied between the two subsystems $L$ and $R$. The $LR$ coupling unwinds the complex winding of the operator size distribution; it winds the size distribution in the opposite direction, accurately reversing the winding direction. The thermal operator, $P$ from the left side, will be exactly mapped to its right side. We obtain a counterclockwise size distribution corresponding to a thermal operator P inserted on the other boundary (the right side $R$) at a positive time \cite{Brown}. The team of researchers stresses: "We explicitly show size-winding of thermal operators near the scrambling time for the SYK model, and we conjecture that the phenomenon can also be found in other holographic systems" \cite{Nezami}.

Perfect-size winding provides a necessary condition for traversable wormhole behavior. It occurs in the ground state, the lowest possible energy where the temperature is zero (low temperature through the wormhole). The ground state of a pair of coupled SYK models is close to the TFD state.

The researchers expect systems with a holographic dual to exhibit perfect size winding \cite{Nezami}. In other words, the SYK model is dual to a traversable wormhole only in the low-temperature regime, and it exhibits perfect size winding in the low-temperature limit. But this applies to large $N$ Majorana fermions interacting with large $q$ other Majorana fermions (teleportation of $q$ fermions). 

But one cannot simulate such a dense SYK model on a noisy quantum devide. So Jafferis and his team asked: what is the most simplified Hamiltonian that preserves the gravitational physics of the dense SYK model?
How many qubits do we need to simulate this Hamiltonian on a quantum device? \emph{Sparsification} reduced the complexity of the dense SYK model. It was shown via classical techniques that an $N = 10$ SYK model was sufficient to produce the traversable wormhole behavior, and its ground state was close to a TFD state. But the team employed classical learning techniques to find a sparser and smaller version of the $N = 10$ SYK model because this model requires too many gates (millions of gates) to be feasible on a Google Sycamore chip. The team constructed an analog of a neural network, backpropagated over the Hamiltonian coefficients, and applied regularization, interpreting the Hamiltonian coefficients as neural network weights.   

More specifically, simulating a many-body Hamiltonian requires many quantum gates. Whereas larger Hamiltonians may provide a stronger teleportation signal, more gates at current quantum hardware fidelity further attenuate the signal. A simplified learned Hamiltonian was constructed. Researchers, therefore, decided to restrict their attention to the smallest sparsified model with gravitational properties.  
The techniques from machine learning and a kind of approximation called \emph{Trotterization} were applied to optimize the procedure. The techniques were performed on a classical computer, and teleportation with $N = 10$ was simulated on a classical computer. 
First, researchers verified that an $N = 10$ SYK model exhibits wormhole-like behavior. They subsequently used machine learning to reproduce the teleportation behavior of the $N = 10$ SYK model with only a few Hamiltonian terms [see equation (\ref{Eq10}), section \ref{10}]. 

The sparsification procedure reduced the $N = 10$ SYK Hamiltonian with $210$ terms to a large population of candidate sparse Hamiltonians thought to preserve the gravitational properties of the original model. Further simulations were run, and the smallest possible model that could exhibit the gravitational phenomena was chosen. We start with a Hamiltonian dual to a two-dimensional AdS (AdS2) space and choose $q = 4$:

\begin{equation} \label{Eq7}
H_L= \sum_{1\leq i<j <k <l\leq N} J_{ijkl}\psi_{L }^{i} \psi_{L }^{j}\psi_{L }^{k} \psi_{L }^{l} , H_R= \sum_{1\leq i<j <k <l\leq N} J_{ijkl}\psi_{R }^{i} \psi_{R }^{j}\psi_{R }^{k} \psi_{R }^{l}.   	  
\end{equation}

\begin{equation} \label{Eq2}
H_{L,R}=-0.36\psi^1\psi^2\psi^4\psi^5+0.19\psi^1\psi^3\psi^4\psi^7 	  
\end{equation}

$-0.71\psi^1\psi^3\psi^5\psi^6+0.22\psi^2\psi^3\psi^4\psi^6+0.49\psi^2\psi^3\psi^5\psi^7,$ 
\vspace{3mm} %3mm vertical space

\noindent with $N = 7$, with only five terms was selected.

As the equations (\ref{Eq7}) and (\ref{Eq2}) show, the Hamiltonian is doubled to give left $H_L$ and right $H_R$ Hamiltonians with $7$ Majorana fermions on each side. Each side simulates the SYK model. The wormhole teleportation protocol also introduces a pair of entangled qubits, i.e., a reference qubit entangled with the injected qubit. Thus the total circuit has $9$ qubits \cite{Jafferis}. 

The learned Hamiltonian equation \ref{Eq2} scrambles and thermalizes similarly to the original $N = 10$ SYK model as characterized by two-point and four-point functions (the two-point function indicates thermalization time while the four-point function indicates scrambling time). The transmitted fermions must thermalize and scramble in the left side $H_L$, as defined by the decay of two-point correlators and OTOCs (see section \ref{1} for definition). Researchers plotted the average of each correlator over Majorana operators, indicating a correspondence between the $N = 10$ SYK and learned Hamiltonian curves \cite{Jafferis}, \cite{Jafferis2}.

Many-body quantum systems thermalize. Thermalization is intimately related to information scrambling and quantum chaos. Large-$N$ SYK model satisfies a \emph{scrambling time} (just before the onset of the chaotic behavior) much bigger than the thermalization time. But the learned Hamiltonian equation (\ref{Eq2}) satisfies a scrambling time approximately equal to the thermalization time, which is compatible with wormhole-like teleportation. 
Before the weak coupling interaction between the $L$ and $R$ subsystems is applied, the transmitted qubits must thermalize and scramble in $L$. Applying the interaction must ensure that time evolution causes the qubits to unthermalize and unscramble on the $R$ subsystem \cite{Jafferis1}.

To find the ground state of the learned Hamiltonian (equation \ref{Eq2}), the researchers apply a hybrid classical-quantum algorithm called \emph{Variational Quantum Eigensolver} (VQE). They apply the VQE to the following Hamiltonian: 

\begin{equation} \label{Eq6}
H_{TFD}= H_L + H_R + i\mu V, 	  
\end{equation}

\noindent where $V$ is the interaction term:

\begin{equation} \label{Eq9}
V=\sum_j \psi{^j_L}\psi{^j_R}, 	  
\end{equation}

\noindent and $\mu$ represents the coupling interaction. It is found that the ground state of $H_{TFD}$ is approximately the TFD state.

Equation (\ref{Eq6}) looks like the wormhole equivalent, the \emph{eternal traversable wormhole Hamiltonian}, which is denoted by $H_{tot}$, due to the ground state of the coupled Hamiltonian:

\begin{equation} \label{Eq5}
H_{tot}= H_L + H_R + i\mu' V. 	  
\end{equation}

The experiment simulated Majorana's fermions. That said, the team of researchers is speaking of the Majorana SYK model with $N$ fermions with which they produce evidence of gravitational physics in the sparsified SYK system: "To encode $7$ Majorana fermions on the left system and $7$ Majorana fermions on the right system, we require $7$ qubits (two fermions per qubit)" \cite{Jafferis}. It is also written: "We assume that the total number of qubits (or fermions) on each side is $n$, and the number of message qubits (or fermions) that are transmitted by the state transfer or operator transfer protocols is $m$" \cite{Nezami}. 

At about the same time, Susskind and a team of researchers were working on what seemed like a bigger project, a sparse SYK model that recovers the global physics of ordinary SYK models. In particular, at low temperatures, their model exhibits a gravitational sector that is maximally chaotic. The sparsity of the model, so writes the team, "considerably reduces the cost of quantum simulation algorithms". This, claims the team, makes their sparse SYK model "the most efficient currently known route to simulate a holographic model of quantum gravity". The researchers add: "On a practical level, sparse systems typically admit much more efficient computer simulations—both classical and quantum. By significantly reducing the resources needed to simulate black holes in holographic quantum gravity models, these results bring us closer to the goal of studying 'quantum gravity in the lab'" \cite{Xu}. 

\section{The quantum circuit and the gravity picture} \label{5}

Researchers have created the following quantum circuit for wormhole teleportation (based on \cite{Brown} and \cite{Jafferis2}):

1) Two identical copies of the quantum system are prepared: a system of $7$ qubits on the left (side $L$) and a system of $7$ qubits on the right (side $R$). The two subsystems are prepared in the TFD state at $t = 0$. 

2) We evolve all the qubits on the side $L$ (register $L$) backward in time by acting with the inverse of the time-evolution operator ($\exp^{+iHt}$) (performed to first order with a single Trotter step. The Trotterization does not introduce significant error).

3) Register $P$ holds a reference qubit $Q$ entangled with a qubit $P$ on register $Q$, both are inserted into the wormhole to check whether the protocol works. A qubit $P$ is then injected at the time $t= -t_0$, i.e., a SWAP gate is used for swapping the qubit $P$, the message, to the side $L$.

4) Now we evolve subsystem $L$ (register $L$) forward in time using the time-evolution operator ($\exp^{-iHt}$) (performed to first order with a single Trotter step).
As a result, $P$ is scrambled among (entangled with) the $7$ qubits on the subsystem $L$ (the carrier qubits). 

5) We now weakly couple side $L$ to side $R$ (at $t = 0$), applying a coupling operator $\exp^{i\mu V}$ (performed to first order with a single Trotter step), where $V$ the coupling operator is defined in equation (\ref{Eq9}). The coupling is applied suddenly: All the $7$ qubits on side $L$ are now coupled to the $7$ qubits on side $R$.  

6) We now evolve side $R$ forward in time using the time-evolution operator ($\exp^{-iHt}$). Side $R$ is subsequently measured (you read out the register $T$). The qubit $P$ (the message) reappears unscrambled, arrives at arrival time $t=t_1$ unscathed at $R$, and there is no need to decode it (a final SWAP gate: extract qubit $P$ from $R$). The message is teleported if the register $P$ is entangled with the register $T$. 

Researchers ran the quantum circuit several times on the Sycamore chip. 

There are two mechanisms of transmission with the wormhole circuit \cite{Brown}: 
1) The low-temperature teleportation: If $\mu < 0$, the qubit $P$ experiences a time advance and is rescued on the side $R$. This is wormhole teleportation. 2) Conversely, when $\mu > 0$ the qubit is entangled with the qubits of side $L$ but is not unscrambled and its destiny is oblivion. This is called scrambling teleportation.

To measure the entanglement of the qubits, one computes the mutual information $I_{PT}(t)$:

\begin{equation} \label{Eq8}
I_{PT}(t) = S_{P}(t) + S_{T}(t) - S_{PT}(t),    	  
\end{equation}

\noindent where $S$ is the von Neumann entropy. The entropy measure was computed in terms of Rényi entropy in the experiment. 

When $\mu < 0$, a peak in the signal $I_{PT}(t)$ is reached ($t$ is of the order of the scrambling time), indicating quantum teleportation in the time window when the wormhole is traversable. 

Jafferis and his team of researchers have confirmed that the learned Hamiltonian's behavior is consistent with the original $N = 10$ SYK model by $I_{PT}(t)$ for $\mu = -12$ and $\mu = +12$. They show this by a similarity in the curves of the two models. The similarity is demonstrated for the peak positions of the $N = 10$ SYK model and the learned Hamiltonian [equation \ref{Eq2}].  

Jafferis and the team of researchers present graphs with curves that show that their coarse-grained SYK model preserves key properties of the traversable wormhole physics: perfect size winding, coupling interaction on either side of the wormhole that is consistent with a negative energy shock wave, a Shapiro time delay, causal time-order of signals emerging from the wormhole, and scrambling and unscrampling \cite{Jafferis}. 

Besides tossing a single qubit on the left side and sending it to the right side, another qubit is thrown on the right side and sent from right to left. The result is time-ordered teleportation, interpreted as a demonstration of gravitational teleportation. At time $-t_0$, a qubit $Q$ is swapped into $L$. Simultaneously, a qubit $R$ is swapped into $R$. At the time $t_1$, the team of researchers performs a measurement and compares the two processes. 
Qubits inserted earlier will emerge earlier, and qubits inserted later should pop out later, so the causal time ordering indicates gravitational interpretation and is extremely important.  
But if you send one qubit into $L$ and another into $R$, they should meet in the middle. But it causes a little delay. It was found that a qubit injected to $R$ would scatter with a qubit injected to $L$, causing a \emph{Shapiro time delay}. It is also demonstrated that in the high-temperature regime, non-gravitational teleportation occurs, and there is no size winding \cite{Jafferis}, \cite{Zlokapa1}. 

Working with collaborators from Caltech, Fermilab, and Harvard, the quantum system was subjected to numerous tests to determine if it showed quantum gravitational behavior. The above signatures were verified on classical computers, so claims Jafferis and the team of researchers, confirming that the quantum system dynamics were consistent with a quantum gravity interpretation and the holographic principle \cite{Zlokapa}.

According to Occam’s razor, the simplest explanation for the above mechanism is teleportation-by-size, i.e., holographic teleportation through a "baby" wormhole. Thus, a message has been teleported in the gravity picture through a semi-classical holographic traversable wormhole \cite{Brown}. Holographically, the above coupled $LR$ quantum system (the CFT side of the duality) is dual to an eternal wormhole that connects the two sides of the eternal black hole (the AdS side of the duality). 

The $LR$ coupling, applied between the two sides of the wormhole, renders the wormhole traversable; if $\mu < 0$, the coupling operator generates a negative energy shockwave in bulk, modifying the geometry of the wormhole and allowing traversability. The traversing qubit experiences a Shapiro time advance on contact with the pulse of negative energy shockwave, causing it to emerge on the other $R$ side at $t = t_1$.
The ANEC requires the negative energy shockwave between the two sides $L$ and $R$ to traverse the semi-classical wormhole. 

When $\mu > 0$, the coupling generates a positive energy shockwave; the qubit falls into the singularity. 

Jafferis and the team of researchers write: "We observe increased teleportation when the interaction introduces a negative energy shockwave rather than a positive one. The asymmetric signature is consistent with the physical interpretation that the qubit underwent teleportation through the wormhole" \cite{Jafferis}. 

\section{Some philosophical remarks in memory of Ian Hacking} \label{8}

Ian Hacking, the great Canadian philosopher of science, passed away on May 10th. One of the basic ideas in Hacking’s philosophy of experimentation is laboratory science, defined as follows: “The laboratory sciences use apparatus in isolation to interfere with the course of that aspect of nature that is under study”. In laboratory science, we implement manipulation. Hacking defines “the materiel of the experiment”, which includes three parts, each associated with a set of instruments: certain devices prepare a target; then there is an apparatus used to somehow interfere with (manipulate) the target. Finally, there is a detector that measures the result of the interference or modification of the target. Hacking, a scientific realist (entity realist),  invented the famous slogan: “If you can spray them, then they are real”. Although subatomic particles are unobservable, you can still "spray them” and use them as tools for other experiments. When Hacking speaks about unobservable entities, particularly electrons, he states that the experimenter manipulates an entity (e.g., electrons) and uses it to experiment on something else \cite{Hacking1}.  

According to Hacking's above theory, in the new experiment performed with the Sycamore chip, physicists first choose the best $9$ qubits. They interfere with these superconducting qubits or transmons, constituting the tabletop experiment. Finally, teleportation behavior is detected in the laboratory's quantum setup. 

I will describe the elements of the chip relevant to the discussion here. The wormhole experiment was realized with superconducting qubits on Google's $72$-qubit quantum processor, Sycamore. The Sycamore chip is cooled to dilution refrigerator temperatures. When cooled to cryogenic temperatures, the superconducting circuits behave as quantum mechanical oscillators, i.e., superconducting artificial atoms. Superconducting qubits refer to these artificial atoms' ground state and first-excited state. These two-lowest states form an effective two-level system, a quantum bit of information used as the qubit. But it is very difficult to confine the quantum two-level system to the subspace of just two levels.   
The Sycamore comprises superconducting qubits called \emph{transmons} (transmission-line shunted plasma oscillation qubits). The transmon is  closely related to the charge qubits or Cooper–Pair–Box (CPB) (Cooper pairs tunneling in a Josephson junction). The transmons comprise Josephson junctions, inductors, capacitors, and superconducting interconnects. 
The transmon fixes the CPB's weakness, and compared to the CPB, it greatly reduces charge noise sensitivity in the qubit \cite{Koch}, \cite{Oliver}. Other types of noise are discussed in section \ref{9}.
Although according to Hacking, we manipulate "natural" subatomic particles and use them to experiment on something else. When simulating on quantum computers, we interfere with both "natural" particles and "artificial particles" (qubits).  

Hacking defines two kinds of models:  

1. Models of the phenomena: The word model sometimes means a material model built in the laboratory, i.e., a laboratory simulation of an experiment. For instance, scientists may think more clearly about a certain phenomenon if they build a desktop model (from wires, wood, plastic, glue, pulleys, levers, ball bearings, weights, lasers, etc.). They can even get the right inputs and outputs from the model. The model may assist in better understanding the phenomenon and suggest new improvements to the experiments; however, according to Hacking, “it is not a literal picture of how things really are” \cite{Hacking1}, \cite{Hacking2}, \cite{Hacking3}, \cite{Hacking4}. 

The same holds for the wormhole experiment in the lab; the model involves the creation of phenomena that do not exist as a pure phenomenon to which theory answers, and therefore according to Hacking, no one suggests that this model is, in fact, a traversable wormhole; the laboratory wormhole does not exist as a pure phenomenon.

2. Models of the theory: according to Hacking, models in physics are something “you hold in your head rather than your hands”. Theories are too complex, and we simplify them by creating theoretical models that approximate world representations. Nancy Cartwright holds that models in physics are not true in the real world \cite{Cartwright}. According to Hacking, models are intermediate between the phenomena and the theories, i.e., they move some aspects of real phenomena to the theories that explain them and connect them. While doing so, they simplify mathematical structures and function as approximations \cite{Hacking1}. 

According to Hacking’s definition of laboratory science and the above classification of models, the laboratory wormhole seems weakly related to a real gravitational object. While physicists call it a wormhole, from the entity realist’s point of view, this result has nothing to do with wormholes in the lab. Scott Aaronson and others have adopted this view.\footnote{S. Aaronson, "Google’s Sycamore chip: no wormholes, no superfast classical simulation either." \emph{Shtetl-Optimized} December 6, 2022.}

We usually proceed from the success of an experiment to the conclusion that our explanation is likely to be approximately true or true. We think that if an explanation is the best among the competing explanations of the experiment, then it is probably true. But it should be stressed that the fit between the \emph{simplified} SYK model and the explanation in terms of an emergent wormhole does not mean that the latter explanation is \emph{literally true}.  Neither does it mean that holographic wormholes \emph{exist} or they are \emph{real}. What is meant by saying that this explanation is the simplest among the other hypotheses is mainly that it is the best fit for the experimental setup and that holographic teleportation fits the teleportation mechanism at the basis of the said experiment.

The point is that according to the ER = EPR hypothesis, the gravity picture is equivalent to the quantum information picture, and "The traversable
wormhole expressed as a quantum circuit, equivalent to the gravitational
picture in the semiclassical limit of infinite qubits" \cite{Jafferis}. But although the analogy between the experimental setup and the emergent geometry is suggestive, it does not follow from the experiment that the wormhole gravitational picture is real. We can only say that teleportation-by-size is the hypothesis that explains the experiment best. This is so even if it explains the evidence. "Truth" requires a step beyond the judgment that the holographic wormhole hypothesis fits the experimental setup and the data and is better than all its rivals.

\section{Quantum gravity in the lab}\label{9}

Researchers argue: "The ‘quantum gravity in the lab’ program does not need to wait for large error-corrected quantum computers. Progress can be made even in the Noisy Intermediate-Scale Quantum (NISQ) era" \cite{Nezami}.

The information does not vanish for very low temperatures and chaotic perturbations do not destroy the original entanglement between the two registers, $P$ and $T$ (see section \ref{5}). How is this possible? Although there is scrambling and quantum chaotic behavior, the weak coupling interaction between $L$ and $R$ entangles $L$ and $R$, and the qubit $P$ is unscrambled. This is perfect size winding, which causes teleportation around the scrambling time. In the perfect size winding scrambling protocol followed by unscrambling, the teleported qubit is highly error-protected \cite{Gao}. 

But as is well known, quantum computers are prone to many errors, and the Sycamore quantum device has a large error rate. Certain people believe that slow and steady wins the race \cite{Kalai} because the chip might not be very reliable. In this state of affairs, "If, at any point in time, a small error occurs, the chaotic dynamics will not undo themselves, and the particle will not make it through the wormhole" \cite{Zlokapa}.       

At large times, a small perturbation can destroy the correlations between the two sides $L$ and $R$ of the quantum system that would otherwise exist without the perturbation. Although the qubits of the Sycamore processor are cooled down to cryogenic temperatures and are held in an ultra-high vacuum chamber, the entangled qubits can decohere quickly due to interaction (entanglement) with the environment (incoherent errors). Jafferis and his team of researchers write: "In general, errors can include coherent errors [crosstalk errors and qubit phase] and incoherent sources of noise; in simulations, we assume fully incoherent errors and observe agreement with experimental data" \cite{Jafferis}. 
 
An appropriate ansatz can mitigate coherent errors for only a small number of qubits ($18$) on the Sycamore quantum device \cite{Niu}.  
Leaders of the "quantum gravity in the Lab" program, Brown, Susskind, and their team, show that "with some caveats, we can use a finite fraction of the fermions" \cite{Nezami}. So to reduce the coherent errors, "the total circuit has $9$ qubits". Recall that in practice, only $7$ qubits were used to simulate a "wormhole-like teleportation". The other two qubits served as the teleported qubits (see sections \ref{3} and \ref{5}). 
Using machine learning, the researchers made the quantum model simple enough to preserve the key gravitational properties to realize it with a circuit with $164$ two-qubit gates. A more complex model would increase the number of gates and, consequently, the error rate. Researchers performed XEB calibration and calibrated readout errors over the $9$ qubits region of the Sycamore. Although the Sycamore chip has $72$ qubits, the team tried to find the least noisy ones on the chip and apply the calibration tools to these $9$ qubits. The chip's errors and noise attenuated the teleportation signal, but the team reports that they got a teleportation signal on the $9$ noisy device \cite{Jafferis}, \cite{Zlokapa1}. 

\section{A fly in the ointment} \label{10}

A few months after the publication of the \emph{Nature} paper, three researchers led by  Norman Yao published a comment that pokes holes in Jafferis et al.'s $5$ term commuting learned Hamiltonian [equation (\ref{Eq2})]. Yao and his team asked: is equation (\ref{Eq2}) consistent with the gravitational dynamics (a qubit emerging from a traversable wormhole) of the original SYK model? 

They found a lacuna in equation (\ref{Eq2}), claiming that it neither recovers the global physics of ordinary SYK models nor fully captures the features of gravitational physics \cite{Korbin}: 

1. "The teleportation signal only resembles the SYK model for Majorana operators used in the machine-learning training" \cite{Korbin}. 
The machine-learning procedure trains on teleportation involving two specific fermions $\psi^1$ and $\psi^2$. For instance, Jafferis and his team numerically evaluate the interaction (the inverse of the time evolution operator $\exp^{iHt}$) on the two qubits $\psi^1$ and $\psi^2$. The results show an asymmetry for opposite signs of $\mu$ ($\mu=12$, scrambling teleportation, and $\mu=-12$, wormhole teleportation), demonstrating qualitative agreement of equation (\ref{Eq2}) with numerical simulation of the $N = 10$ SYK model \cite{Jafferis}.  

In other words, argue Yao and his team, the teleportation signal (that is consistent with a negative energy shockwave, causal time-ordering of teleported signals, and a Shapiro time delay) and perfect size winding are characterized only for fermions $\psi^1$ and $\psi^2$ that were trained and not for others that were not involved in the training. For this pair of fermions, there is good agreement between equation (\ref{Eq2}) and the $N = 10$ SYK model. But, say Yao and his team, we do not observe these characteristics when the teleportation protocol is performed with other fermions not involved in the training procedure \cite{Korbin}. 
Korbin et al. found a problem with the following untrained fermions $\psi^4$ and $\psi^7$. They "observe that fermions $\psi^4$ and $\psi^7$ have poor size winding at $t_0\approx 2.8$" (the time of the wormhole teleportation signal or injection time). In other words, size winding and teleportation behavior with $\mu = -12$ and at $t_0\approx 2.8$ is only observed for the trained Majorana fermions $\psi^1$ and $\psi^2$ \cite{Korbin}. 

2. Equation (\ref{Eq2}) has all commuting terms. But there is a difference in the structure of time-evolved operators between fully-commuting models, i.e., equation (\ref{Eq2}), and non-commuting models, the original SYK model. Korbin et al. examine two alternatives to equation (\ref{Eq2}): 

1) In Jafferis et al.'s \emph{Nature} paper, it is shown that the machine-learning procedure that produced equation (\ref{Eq2}) produces a $6$ term Hamiltonian, which, unlike equation (\ref{Eq2}), does not have all commuting terms  \cite{Jafferis}:

\vspace{1mm} %1mm vertical space
\begin{equation} \label{Eq3}
H_{L,R}=-0.35\psi^1\psi^2\psi^3\psi^6+0.11\psi^1\psi^2\psi^3\psi^8 	  
\end{equation}

$-0.17\psi^1\psi^2\psi^4\psi^7-0.67\psi^1\psi^3\psi^5\psi^7+0.38\psi^2\psi^3\psi^6\psi^7-0.05\psi^2\psi^5\psi^6\psi^7.$ 
\vspace{5mm} %5mm vertical space

\noindent Equation (\ref{Eq3}) is weakly perturbed from a fully commuting Hamiltonian.

Yao and his team found a flaw in this Hamiltonian. They show that the main observations regarding equation (\ref{Eq2}) also apply to equation (\ref{Eq3}). 

In particular, the teleportation signal does not resemble the $N = 10$ SYK model for untrained operators. The size winding behavior resembles that of equation (\ref{Eq2}). Further, unlike the $N = 10$ SYK model, within the timescale on which $\psi^1$ and $\psi^2$ were trained, the teleportation signal for equation (\ref{Eq3}) exhibits an initial peak, followed by significant revivals as a function of time. 

The two-point function exhibits decay for both the SYK model and equation (\ref{Eq2}) (see section \ref{3}). For the SYK model, this decay is consistent with thermalization. However, for equation (\ref{Eq2}), the individual two-point correlators exhibit strong revivals as a function of time. This behavior "indicates that the training procedure was not fully successful; such disagreement is not shown or commented on in" the \emph{Nature} paper, say Yao and his team. Thus, the agreement between equation (\ref{Eq2}) and the SYK model is an artifact of averaging over the two-point and four-point correlation functions \cite{Korbin}.

2) In their \emph{Nature} paper, Jafferis et al. use an alternate machine-learning procedure to produce another Hamiltonian with $N=10$ and $8$ terms. This Hamiltonian is designed to maximize the difference in the teleportation
signal between $-\mu$ (wormhole teleportation) and $+\mu$  (scrambling teleportation) \cite{Jafferis}:
\vspace{3mm} %3mm vertical space

\begin{equation} \label{Eq10}
H_{L,R}=0.60\psi^1\psi^3\psi^4\psi^5+0.72\psi^1\psi^3\psi^5\psi^6 	  
\end{equation}

$+0.49\psi^1\psi^5\psi^6\psi^9+0.49\psi^1\psi^5\psi^7\psi^8$
\vspace{1mm} %1mm vertical space

$+0.64\psi^2\psi^4\psi^8\psi^{10}-0.75\psi^2\psi^5\psi^7\psi^8+0.58\psi^2\psi^5\psi^7\psi^{10}-0.53\psi^2\psi^7\psi^8\psi^{10}.$ 
\vspace{3mm} %3mm vertical space

Jafferis and his team write that since the number of gates required to perform Trotterization scales linearly with the number of terms, the gate count to implement the Hamiltonian must at least double. Circuit fidelity exponentially decays with the number of gates and qubits (and the experimentally measured fidelity was already below $1/2$ of the noiseless fidelity). Consequently, Jafferis and his team conclude that equation (\ref{Eq10}) "cannot provide a stronger teleportation signal when experimentally measured", and does not exhibit perfect size winding but a slightly damped one \cite{Jafferis}. 

Referring to equation (\ref{Eq10}), Yao and his team again find a problem with small-size fully-commuting Hamiltonians with a few terms. They comment that this Hamiltonian is non-commuting and exhibits clearer signatures of thermalization at a long time scale $t\approx 30$. Indeed, the teleportation signal for this Hamiltonian exhibits a single peak structure for nearly all operators, and it does not exhibit perfect-size winding. 
The perfect-size winding is, in fact, a generic feature of fully commuting Hamiltonians with only a few terms. But perfect-size winding does not persist in fully-commuting larger and non-commuting larger systems. Yao and his team conclude that perfect-size winding in small-size fully-commuting Hamiltonians with only a few terms is probably a side effect. It does not mean equation (\ref{Eq2}) is holographically dual to gravity \cite{Korbin}. 

Perfect-size winding, however, is necessary for a traversable holographic wormhole. Yao and his team argue that the perfect-size winding reported in the \emph{Nature} paper relies on the small size of the system and equation (\ref{Eq2}). Small-size, fully-commuting Hamiltonians do not thermalize but generally exhibit perfect-size winding. In contrast, the opposite is true for larger or non-commuting systems.
Thus, unlike the $N = 10$ SYK model, equation (\ref{Eq2}) does not thermalize, and the agreement in the thermalization behavior between the two is an artifact \cite{Korbin}.

Korbin et al. summarize their argument: "The authors’ machine-learning procedure may have introduced a bias among the trained operators [...] The fact that the perfect-size winding observed in \cite{Jafferis} seems reliant on small-size, fully-commuting models—which defy other features of holography such as
thermalization, complexity, and chaos — raises the question of whether the observed perfect-size winding is indeed connected to gravitational physics in a
substantive manner". In other words, the perfect size winding ansatz may not be correct in the learned Hamiltonian (for $N =7$) because a bias was imposed on the machine-learning procedure \cite{Korbin}. 

One month later, Jafferis and his team uploaded a comment in which (as expected) they found a way to save their learned Hamiltonian equation (\ref{Eq2}). Jafferis et al. found an elegant way out. They realized they could easily rebut the main arguments advanced by Korbin et al.:  

First, the commuting structure of equation (\ref{Eq2}) is unrelated to size winding properties and dynamics. Equation (\ref{Eq2}) was trained with $t_0 = 2.8$ and $\mu=-12$. The following behavior was found at the time of teleportation: two Majorana operators $\psi^1$ and $\psi^2$ showed fast operator growth (in the gravitational picture, we would say that they traverse the wormhole), while the Majorana operators $\psi^4$ and $\psi^7$ were the slowest to show operator growth. It was argued in the comment that the single-sided Hamiltonian $H_L$ equation (\ref{Eq2}) has commuting terms and does not thermalize at later times after teleportation due to the recurrence of the commuting terms. 
Jafferis et al. found a way, in the context of the eternal traversable wormhole Hamiltonian $H_{tot}$ to rebut this claim: evolving under $H_{tot}$, equation (\ref{Eq5}), from the \emph{Nature} paper, shows that equation (\ref{Eq5}) exhibits operator growth and thermalizes at high temperatures after teleportation. 
A single Trotter step of time evolution under this Hamiltonian is equivalent to the teleportation quantum circuit of the \emph{Nature} paper described in section \ref{5}. The gravitational interpretation of thermalization rates is that each fermion corresponds to a different mass. Inserting a fermion onto the TFD and time-evolving under a single-sided Hamiltonian $H_L$ should result in operator growth, with lighter fermions growing more slowly. The wave packet of a lighter fermion is more spread out, and its two-point function decays more slowly. Jafferis et al. conclude that the two-point function of equation (\ref{Eq5}) decays for all fermions, indicating thermalization. Thus, "We see that the fermions that exhibit the slowest operator growth in $H_L$ are precisely the same fermions that thermalize the slowest in $H_{tot}$", equation (\ref{Eq5}) \cite{Jafferis1}. 

Jafferis et al. found an explanation that allowed them to avoid the problem pointed out by Korbin et al. They added that in equation (\ref{Eq5}), all fermions show size winding between $2\gtrsim t \lesssim 5$. At the same time, in Korbin et al.'s comment, what is claimed as an artifact is only analyzing size winding at $t_0 \approx 2.8$. Jafferis et al. found examining a counterfactual scenario that leads to meaningful gravitational behavior advantageous. They show that at later times $5\gtrsim t \lesssim 10$, wormhole teleportation persists despite the addition of a strong non-commuting perturbation, see [equation (\ref{Eq4})] below. Although a noncommuting Hamiltonian now governs the late-time dynamics, the system's behavior during teleportation at $t_0 \approx 2.8$ is unchanged and remains consistent with the expected gravitational signature \cite{Jafferis1}. 

Yao and his team claim that equation (\ref{Eq2}) "is biased to have good size winding only on the two fermions implemented in the experiment" \cite{Korbin}. Jafferis and his team dismiss this claim, and they find a very interesting physical justification for this: fermions that thermalize more slowly in the eternal traversable wormhole Hamiltonian [equation (\ref{Eq5})] achieve good-size winding at later times. This is again associated with different masses across fermions: lighter fermions thermalize more slowly and take longer to traverse the wormhole. Indeed, fermions $\psi^4$ and $\psi^7$ are the slowest to thermalize. Consequently, the microscopic mechanism of size winding should occur later, and these fermions achieve size winding at slightly later times. Those fermions identified in Korbin et al.'s comment to have poor size winding at $t_0 \approx 2.8$, say Jafferis and his team, also thermalize more slowly and experience slower operator growth; they do so later at $t_0 \approx 4$ instead of $t_0 \approx 2.8$ - the time assumed in the comment. "This is consistent with interpreting them as taking longer to traverse the wormhole". The bottom line is that eventually, all fermions achieve size winding \cite{Jafferis1}.

Second, a large non-commuting term (perturbation) $H_1$ is provided for equation (\ref{Eq2}). Equation (\ref{Eq2}) is called $H_0$. I will stick to this term. Thus the perturbed Hamiltonian is:

\begin{equation} \label{Eq4}
H_0 + H_1 = H_0+0.3\psi^1\psi^2\psi^3\psi^5, 	  
\end{equation}

\noindent which produces a behavior similar to the SYK model. This Hamiltonian thermalizes at later times. Still, it has similar size winding and teleportation behavior with $\mu = -12$ and at $t_0 \approx 2.8$.  

Recall that Korbin et al. found a problem with small-size fully-commuting Hamiltonians with only a few terms, arguing that perfect-size winding in those systems is probably a side effect and does not signify a holographically
dual to gravity \cite{Korbin}.

Jafferis and his team found a simple way to clarify this problem, arguing that equation (\ref{Eq4}) shows a size winding sufficient to produce an asymmetric teleportation signal. The non-commuting perturbation $H_1$ does not significantly affect the physics during teleportation, suggesting the commuting structure is irrelevant to the presence of gravitational physics. Moreover, like the $N = 10$ SYK model, the two fermions traversing the wormhole, $\psi^1$ and $\psi^2$, do not exhibit significant revivals as a function of time after the initial peak \cite{Jafferis1}. 

Finally, Jafferis and his team believe that the presence of size winding in Hamiltonians, such as equation (\ref{Eq3}), explicitly constructed to be similar to equation (\ref{Eq2}), is unsurprising. Still, it is irrelevant in determining a gravitational interpretation of equation (\ref{Eq2}) \cite{Jafferis1}.

The reaction of people to Korbin et al.'s paper and the whole story of the traversable wormhole and the learned Hamiltonian was a typical Popperian reaction: we have here a testable model, the learned Hamiltonian equation (\ref{Eq2}), that was found to be problematic but is still upheld by Jafferis and his team, who reinterpret it to escape refutation. Karl Popper argues that such a procedure is always possible. Still, it rescues the model (the learned Hamiltonian) from refutation at the price of destroying or at least lowering its scientific status. Scientists find it hard to give up their hypotheses. But even Popper warned scientists not to give up their hypotheses too easily and not, at any rate, before they have critically examined them. In the face of apparent refutations, scientists who give up their hypotheses too easily will never discover the possibilities inherent in their model. There is room in science for debate and for defense. But a scientist has to conjecture when to stop defending a favorite hypothesis and when to try a new one (\cite{Popper}). 
Although Korbin et al.'s paper found flaws in Jafferis et al.'s learned Hamiltonian, we must wait to see how the scientific community reacts to Jafferis et al.'s original solution that allowed them to avoid the problems. 

\section*{Acknowledgement}

\noindent This work is supported by ERC advanced grant number 834735.


\begin{thebibliography}{35}

\bibitem[1]{Bennet}  C. H. Bennet, G. Brassard, C. Crépeau, R. Jozsa, A. Peres and W. K. Wootters (1993). “Teleporting an unknown quantum state via dual classical and Einstein-Podolsky-Rosen Channels.” \emph{Physical Review Letters} 70, pp. 1895-1899.

\bibitem[2]{Blok} M. S. Blok, V. V. Ramasesh, T. Schuster, K. O’Brien, J. M.
Kreikebaum, D. Dahlen, A. Morvan, B. Yoshida, N. Y.
Yao and I. Siddiqi (2021). "Quantum Information Scrambling in a
Superconducting Qutrit Processor." \emph{Physical Review X} 10, pp. 021010-1-
 021010-21.

\bibitem[3]{Brown} A. R. Brown, H. Gharibyan, S. Leichenauer, H. W. Lin, S. Nezami, G. Salton, L. Susskind, B. Swingle and M. Walter (2019). "Quantum Gravity in the Lab: Teleportation by Size and Traversable Wormholes." \emph{arXiv}: 1911.06314v2  [quant-ph]

\bibitem[4]{Brown2} A. R. Brown, H. Gharibyan, S. Leichenauer, H. W. Lin, S. Nezami, G. Salton, L. Susskind, B. Swingle, and M. Walter (2023). "Quantum Gravity in the Lab. I. Teleportation by Size and Traversable Wormholes." \emph{PRX QUANTUM a Physical Review journal} 4, pp. 010320-1-010320-22.

\bibitem[5]{Cartwright} N. Cartwright (1983). \emph{How the Laws of Physics Lie.} Oxford: The Clarendon Press.

\bibitem[6]{Gao1} P. Gao and D. L. Jafferis and A, C. Wall (2017). "Traversable wormholes via a double trace deformation." \emph{Journal of High Energy Physics volume} 2017, pp. 1-24. 

\bibitem[7]{Gao} P. Gao and D. L. Jafferis (2021). "A traversable wormhole teleportation protocol in the SYK model." \emph{Journal of High Energy Physics} 2021, pp. 1-43.

\bibitem[8]{Hacking1} I. Hacking (1983). \emph{Representing and Intervening: Introductory. Topics in the Philosophy of Natural Science.} Cambridge: Cambridge University Press.

\bibitem[9]{Hacking2} I. Hacking (1988). “On the Stability of the Laboratory Sciences.” \emph{The Journal of Philosophy} 85, pp. 507-514.

\bibitem[10]{Hacking3} I. Hacking (1988). “Philosophers of Experiment.” \emph{PSA: Proceedings of the Biennial Meeting of the Philosophy of Science Association 1988, Volume Two: Symposia and Invited Papers}, pp. 147-156.

\bibitem[11]{Hacking4} I. Hacking (1992). “The Self-Vindication of the Laboratory Sciences.” In: A. Pickering (ed.) \emph{Science as Practice and Culture}. Chicago, IL: The University of Chicago Press, pp. 29-64. 

\bibitem[12]{Harlow} D. Harlow (2016). "Jerusalem Lectures on Black Holes and
Quantum Information." \emph{Reviews of Modern physics}, pp. 015002-1-015002-58.

\bibitem[13]{Hosur} P. Hosur, X-L. Qi, D. A. Roberts, and B. Yoshida, (2016). "Chaos in Quantum Channels," \emph{Journal of High Energy Physics} 2016, pp. 1-48.

\bibitem[14] {Jafferis} D. Jafferis, A. Zlokapa, J. D. Lykken, D. K. Kolchmeyer, S. I. Davis, N. Lauk, H. Neven, and M. Spiropulu (2022). "Traversable wormhole dynamics on a quantum processor." \emph{Nature} 612, pp. 51–55.

\bibitem[15] {Jafferis1} D. Jafferis, A. Zlokapa, J. D. Lykken, D. K. Kolchmeyer, S. I. Davis, N. Lauk, H. Neven, and M. Spiropulu (2023). "Comment on 'Comment on ‘Traversable wormhole dynamics on a quantum processor’.” 	\emph{arXiv:$2303.15423$} [quant-ph], pp. 1-5.

\bibitem[16] {Jafferis2} D. Jafferis (2022). "Emergent Gravitational Dynamics in Quantum Experiments." \emph{Symposium on Quantum Information, Complexity, and the Physical World}, Monday, December 5, 2022 - Princeton University.

\bibitem[17] {Kalai} G. Kalai, Y. Rinott, T. Shoham (2022). "Google's 2019 'Quantum Supremacy' Claims: Data, Documentation, and Discussion." \emph{arXiv}:$2210.12753v2$ [quant-ph], pp. 1-34.

\bibitem[18] {Koch} J. Koch, T. M. Yu, J. Gambetta, A. A. Houck, D. I. Schuster, J. Majer, A. Blais, M. H. Devoret, S. M. Girvin and R. J. Schoelkopf (2007). "Charge-insensitive qubit design derived from the Cooper pair box." \emph{Physical Review A} 76, pp. 042319-1-042319-19.

\bibitem[19] {Korbin} B. Kobrin, T. Schuster, N. Y. Yao (2023). "Comment on 'Traversable wormhole dynamics on a quantum processor'." \emph{arXiv}:$2302.07897$ [quant-ph], pp. 1-9.

\bibitem[20] {Landsman} K. A. Landsman, C. Figgatt, T. Schuster, N. M. Linke, B. Yoshida, N. Y. Yao, and C. Monroe (2019). "Verified quantum information scrambling." \emph{Nature} 567, pp. 61–65.

\bibitem[21] {Maldacena2} J. Maldacena (2003). "Eternal black holes in anti-de Sitter." \emph{Journal of High Energy Physics} 2003, pp. 1-16.

\bibitem[22] {Maldacena1} J. Maldacena and X.-L. Qi (2018) Eternal traversable wormhole, \emph{arXiv}:1804.00491v3 [hep-th], pp. 1-74. 

\bibitem[23] {Maldacena} J. Maldacena, S. H. Shenker and D. Stanford, (2016). "A Bound on Chaos." \emph{Journal of High Energy Physics} (2016), pp. 1-16.

\bibitem[24]{Niu} M. Y. Niu, A. Zlokapa, M. Broughton, S. Boixo, M. Mohseni, V. Smelyanskyi, and H. Neven (2022). "Entangling Quantum Generative Adversarial Networks."\emph{Physical Review Letters} 128.220505.

\bibitem[25]{Nezami} S. Nezami, H. W. Lin, A. R. Brown, H. Gharibyan, S. Leichenauer, G. Salton, L. Susskind, B. Swingle and M. Walter (2022). "Quantum Gravity in the Lab: Teleportation by Size and Traversable Wormholes, Part II." \emph{arXiv}: 2102.01064v1  [quant-ph] 

\bibitem[26]{Nezami2} S. Nezami, H. W. Lin, A. R. Brown, H. Gharibyan, S. Leichenauer, G. Salton, L. Susskind, B. Swingle and M. Walter (2022). "Quantum Gravity in the Lab. II. Teleportation by Size and Traversable Wormholes" Quantum Gravity in the Lab. I. Teleportation by Size and Traversable Wormholes." \emph{PRX QUANTUM a Physical Review journal} 4, pp. 010321-1-010321-33.

\bibitem[27]{Oliver} W. D. Oliver and P. B. Welander (2013). "Materials in superconducting quantum bits." \emph{MRS Bulletin} 38, pp. 816-825.

\bibitem[28]{Popper} K. R. Popper. (1963). \emph{Conjectures and Refutations: The Growth of Scientific Knowledge}. New York: Routledge.  

\bibitem[29]{Susskind1}  L. Susskind (2017). “Dear Qubitizers. GM = QM.” \emph{arXiv}: hep.th/$1708.03040v1$, pp. 1-14.

\bibitem[30]{Susskind}  L. Susskind and Y. Zhao (2017). “Teleportation Through the Wormhole.” \emph{Physical Review D} 98, pp. 046016-1-046016-18.

\bibitem[31]{Sekino}  Y. Sekino and L. Susskind (2008). "Fast scramblers." \emph{Journal of High Energy Physics} 10, pp. 1-14.

\bibitem[32]{Yoshida} B. Yoshida and N. Y. Yao (2019). "Disentangling Scrambling and Decoherence via Quantum Teleportation." \emph{Physical Review X} 9, pp. 011006-1-011006-17.

\bibitem[33]{Xu} S. Xu, L. Susskind, Y. Su, B. Swingle (2020). "A Sparse Model of Quantum Holography." \emph{arXiv}:2008.02303v1 [cond-mat.str-el], pp. 1-55. 

\bibitem[34]{Zlokapa} Zlokapa, A. (2022). "Making a Traversable Wormhole with a Quantum Computer." \emph{Google Research}. 

\bibitem[35]{Zlokapa1} Zlokapa, A. (2023). "Traversable wormhole dynamics on a quantum processor." Lecture. \emph{Harvard CMSA Quantum Matter in Math and Physics}, 24.3.23.

\end{thebibliography}
\end{document}